\documentclass[twocolumn]{aastex61}
\usepackage{amsmath}
\usepackage{footmisc}
\usepackage{color}
\usepackage{multirow}
\usepackage[normalem]{ulem}

\begin{document}

\title{The 1.4 mm core of Centaurus A: First VLBI results with the South Pole Telescope}

\author{Junhan Kim}
\affiliation{Department of Astronomy and Steward Observatory, University of Arizona, 933 N. Cherry Ave., Tucson, AZ 85721, USA}
\email{E-mail: junhankim@email.arizona.edu}

\author{Daniel P. Marrone}
\affiliation{Department of Astronomy and Steward Observatory, University of Arizona, 933 N. Cherry Ave., Tucson, AZ 85721, USA}

\author{Alan L. Roy}
\affiliation{Max-Planck-Institut f{\"u}r Radioastronomie, Auf dem H{\"u}gel, 69, 53121 Bonn, Germany}

\author{Jan Wagner}
\affiliation{Max-Planck-Institut f{\"u}r Radioastronomie, Auf dem H{\"u}gel, 69, 53121 Bonn, Germany}
\affiliation{Korea Astronomy and Space Science Institute (KASI), 776 Daedeokdae-ro, Yuseong-gu, Daejeon 305-348, Republic of Korea}

\author{Keiichi Asada}
\affiliation{Institute of Astronomy and Astrophysics, Academia Sinica, P. O. Box 23-141, Taipei 10617, Taiwan}

\author{Christopher Beaudoin}
\affiliation{Massachusetts Institute of Technology, Haystack Observatory, Route 40, Westford, MA 01886, USA}

\author{Jay Blanchard}
\affiliation{Departamento de Astronom{\'i}a, Universidad de Concepci{\'o}n, Casilla 160, Chile}
\affiliation{Joint Institute for VLBI ERIC, Postbus 2, 7990 AA Dwingeloo, The Netherlands}

\author{John E. Carlstrom}
\affiliation{Kavli Institute for Cosmological Physics, University of Chicago, 5640 South Ellis Avenue, Chicago, IL 60637, USA}
\affiliation{Department of Astronomy and Astrophysics, University of Chicago, 5640 South Ellis Avenue, Chicago, IL 60637, USA}
\affiliation{Department of Physics, University of Chicago, 5640 South Ellis Avenue, Chicago, IL 60637, USA}
\affiliation{Enrico Fermi Institute, University of Chicago, 5640 South Ellis Avenue, Chicago, IL 60637, USA}

\author{Ming-Tang Chen}
\affiliation{Institute of Astronomy and Astrophysics, Academia Sinica, P. O. Box 23-141, Taipei 10617, Taiwan}

\author{Thomas M. Crawford}
\affiliation{Kavli Institute for Cosmological Physics, University of Chicago, 5640 South Ellis Avenue, Chicago, IL 60637, USA}
\affiliation{Department of Astronomy and Astrophysics, University of Chicago, 5640 South Ellis Avenue, Chicago, IL 60637, USA}

\author{Geoffrey B. Crew}
\affiliation{Massachusetts Institute of Technology, Haystack Observatory, Route 40, Westford, MA 01886, USA}

\author{Sheperd S. Doeleman}
\affiliation{Harvard-Smithsonian Center for Astrophysics, 60 Garden Street, Cambridge, MA 02138, USA}

\author{Vincent L. Fish}
\affiliation{Massachusetts Institute of Technology, Haystack Observatory, Route 40, Westford, MA 01886, USA}

\author{Christopher H. Greer}
\affiliation{Department of Astronomy and Steward Observatory, University of Arizona, 933 N. Cherry Ave., Tucson, AZ 85721, USA}

\author{Mark A. Gurwell}
\affiliation{Harvard-Smithsonian Center for Astrophysics, 60 Garden Street, Cambridge, MA 02138, USA}

\author{Jason W. Henning}
\affiliation{Kavli Institute for Cosmological Physics, University of Chicago, 5640 South Ellis Avenue, Chicago, IL 60637, USA}
\affiliation{Department of Astronomy and Astrophysics, University of Chicago, 5640 South Ellis Avenue, Chicago, IL 60637, USA}

\author{Makoto Inoue}
\affiliation{Institute of Astronomy and Astrophysics, Academia Sinica, P. O. Box 23-141, Taipei 10617, Taiwan}

\author{Ryan Keisler}
\affiliation{Dept. of Physics, Stanford University, 382 Via Pueblo Mall, Stanford, CA 94305, USA}
\affiliation{Kavli Institute for Particle Astrophysics and Cosmology, Stanford University, 452 Lomita Mall, Stanford, CA 94305}

\author{Thomas P. Krichbaum}
\affiliation{Max-Planck-Institut f{\"u}r Radioastronomie, Auf dem H{\"u}gel, 69, 53121 Bonn, Germany}

\author{Ru-Sen Lu}
\affiliation{Max-Planck-Institut f{\"u}r Radioastronomie, Auf dem H{\"u}gel, 69, 53121 Bonn, Germany}

\author{Dirk Muders}
\affiliation{Max-Planck-Institut f{\"u}r Radioastronomie, Auf dem H{\"u}gel, 69, 53121 Bonn, Germany}

\author{Cornelia M{\"u}ller}
\affiliation{Department of Astrophysics/IMAPP, Radboud University Nijmegen, PO Box 9010, 6500 GL Nijmegen, The Netherlands}
\affiliation{Max-Planck-Institut f{\"u}r Radioastronomie, Auf dem H{\"u}gel, 69, 53121 Bonn, Germany}

\author{Chi H. Nguyen}
\affiliation{Department of Astronomy and Steward Observatory, University of Arizona, 933 N. Cherry Ave., Tucson, AZ 85721, USA}
\affiliation{Center for Detectors, School of Physics and Astronomy, Rochester Institute of Technology, 1
Lomb Memorial Dr., Rochester NY 14623, USA}

\author{Eduardo Ros}
\affiliation{Max-Planck-Institut f{\"u}r Radioastronomie, Auf dem H{\"u}gel, 69, 53121 Bonn, Germany}
\affiliation{Observatori Astron{\`o}mic, Universitat de Val{\`e}ncia, 46980 Paterna, Val{\`e}ncia, Spain}
\affiliation{Dept. d'Astronomia i Astrof{\'i}sica, Universitat de Val{\`e}ncia, 46100 Burjassot, Val{\`e}ncia, Spain}

\author{Jason SooHoo}
\affiliation{Massachusetts Institute of Technology, Haystack Observatory, Route 40, Westford, MA 01886, USA}

\author{Remo P. J. Tilanus}
\affiliation{Department of Astrophysics/IMAPP, Radboud University Nijmegen, PO Box 9010, 6500 GL Nijmegen, The Netherlands}
\affiliation{Leiden Observatory, Leiden University, P.O. Box 9513, 2300 RA Leiden, The Netherlands}

\author{Michael Titus}
\affiliation{Massachusetts Institute of Technology, Haystack Observatory, Route 40, Westford, MA 01886, USA}

\author{Laura Vertatschitsch}
\affiliation{Harvard-Smithsonian Center for Astrophysics, 60 Garden Street, Cambridge, MA 02138, USA}

\author{Jonathan Weintroub}
\affiliation{Harvard-Smithsonian Center for Astrophysics, 60 Garden Street, Cambridge, MA 02138, USA}

\author{J. Anton Zensus}
\affiliation{Max-Planck-Institut f{\"u}r Radioastronomie, Auf dem H{\"u}gel, 69, 53121 Bonn, Germany}

\begin{abstract}
Centaurus A (Cen A) is a bright radio source associated with the nearby galaxy NGC 5128 where high-resolution radio observations can probe the jet at scales of less than a light-day. The South Pole Telescope (SPT) and the Atacama Pathfinder Experiment (APEX) performed a single-baseline very-long-baseline interferometry (VLBI) observation of Cen A in January 2015 as part of VLBI receiver deployment for the SPT. 
We measure the correlated flux density of Cen A at a wavelength of 1.4~mm on a $\sim$7000 km (5 G$\lambda$) baseline. Ascribing this correlated flux density to the core, and with the use of a contemporaneous short-baseline flux density from a Submillimeter Array observation, we infer a core brightness temperature of $1.4 \times 10^{11}$~K. This is close to the equipartition brightness temperature, where the magnetic and relativistic particle energy densities are equal. Under the assumption of a circular Gaussian core component, we derive an upper limit to the core size $\phi = 34.0 \pm 1.8~\mu\textrm{as}$, corresponding to 120~Schwarzschild radii for a black hole mass of $5.5 \times 10^7 M_{\odot}$.
\end{abstract}

\section{Introduction}
Centaurus A (PKS 1322-428, hereafter Cen A) is the brightest radio source associated with the galaxy NGC 5128 (see \citealt{1998A&ARv...8..237I} for a review) and located at a distance of $3.8\pm0.1~\textrm{Mpc}$ \citep[][and references therein]{2004A&A...413..903R, 2007AJ....133..504K, 2010PASA...27..457H}. It has a prominent double-sided jet and belongs to the Fanaroff-Riley type I class of radio galaxies \citep{1974MNRAS.167P..31F}. Its proximity and brightness make it an especially suitable target for high-angular-resolution observations with very-long-baseline interferometry (VLBI), which can reveal the jet structure as well as the innermost region of the active galactic nucleus \citep[AGN;][]{2017A&ARv..25....4B}.

Although the Very Long Baseline Array (VLBA) has monitored the source from the northern hemisphere \cite[e.g.,][]{2001ApJ...546..210T, 2001AJ....122.1697T} at wavelengths as short as 7 mm \citep{1997ApJ...475L..93K}, observations of Cen A have been mostly limited to longer wavelength VLBI arrays located in the southern hemisphere \cite[e.g.,][]{1998AJ....115..960T, 2011A&A...530L..11M, 2014A&A...569A.115M} due to its low declination of $-43^{\circ}$. Previous observations have achieved angular resolutions of 0.4~mas $\times$ 0.7~mas at 3.6 cm using an array spanning Australia, Chile, and Antarctica \citep{2011A&A...530L..11M} through the Tracking Active Galactic Nuclei with Austral Milliarcsecond Interferometry \citep[TANAMI;][]{2010A&A...519A..45O, 2018A&A...610A...1M} program and 0.6 mas at 6.1 cm with the VLBI Space Observatory Programme \citep[VSOP;][]{2006PASJ...58..211H} satellite. Observations of the source would benefit from the inclusion of southern stations and measurements at short millimeter wavelengths where the radio core could be explored on smaller scales.

VLBI observations of Cen A have probed its morphology in multiple wavelengths from 13 cm to 7 mm \citep{1998AJ....115..960T, 2001AJ....122.1697T, 2006PASJ...58..211H, 2011A&A...530L..11M, 2014A&A...569A.115M}. \cite{1998AJ....115..960T} and \cite{2014A&A...569A.115M} find that there is a compact component within the jet structure. The VLBA data of \cite{1997ApJ...475L..93K} suggests that the observed structure is already dominated by a single component at 7 mm wavelength. At shorter wavelengths, the lower synchrotron opacity should provide access to deeper regions of the stationary core.

The Event Horizon Telescope (EHT) is a VLBI network operating at 1.4 mm, and in the near future at 0.9 mm \citep[e.g.,][]{2008Natur.455...78D, 2009astro2010S..68D, 2012Sci...338..355D}. Most of the EHT stations are located in the northern hemisphere, namely the Submillimeter Array (SMA; \mbox{8 $\times$ 6~m} dishes) and the James Clerk Maxwell Telescope (JCMT; 15~m) in Hawaii, the Submillimeter Telescope (SMT; 10~m) in Arizona, the Institute de Radioastronomie Millim{\'e}trique (IRAM) 30 m telescope in Spain, and the Large Millimeter Telescope (LMT; 50~m) in Mexico. The sensitivity, imaging capability, and north-south extent of the array have recently been improved through the inclusion of southern hemisphere stations: Atacama Pathfinder Experiment (APEX; 12~m) and the Atacama Large Millimeter/submillimeter Array \citep[ALMA; $\sim$37 $\times$ 12 m dishes;][]{2018PASP..130a5002M} in Chile, and the South Pole Telescope \citep[SPT; 10~m;][]{2011PASP..123..568C}. These sites will enable the EHT to provide better imaging of Cen~A at a much shorter wavelength than past VLBI experiments.

Cen A is powered by a supermassive black hole with a mass of $5.5\pm3.0 \times 10^7 M_{\odot}$ \citep{2009MNRAS.394..660C, 2010PASA...27..449N}. The apparent diameter of the black hole event horizon, accounting for its own gravitational lensing, is $\sim$5$R_\textrm{sch}$ where $R_\textrm{sch}$ is the Schwarzschild radius \citep{1973blho.conf..215B}. The corresponding apparent angular size of the event horizon is 1.5 $\mu$as, well below the \mbox{$\sim$20 $\mu$as} resolution of the EHT at 1.4 mm. However, the EHT still provides the resolution to observe the inner region close to the black hole. For example, the 7000~km baseline between the SPT and APEX provides a fringe spacing of \mbox{40 $\mu$as} (\mbox{150 au} at the distance of Cen A at 1.4 mm). This is better angular resolution than any VLBI observation of Cen A published to date.

In this paper, we report results from the VLBI observation of Cen A with the SPT and APEX at 1.4 mm during commissioning observations for the SPT VLBI system. In section \ref{sec:obs}, we describe the observation and the visibility amplitude calibration. In section \ref{sec:analysis}, we present the analysis of the data to infer physical properties of the radio core of Cen A. This single-baseline observation places a lower limit on the brightness temperature of the Cen A core region, and, when used with the zero-baseline flux density measurement, allows us to place an upper limit on the core size.

\section{Observations and Data reduction}
\label{sec:obs}

\subsection{Observations}
\begin{deluxetable*}{cccccccccc}[t]
\tablecolumns{10}
\tablewidth{0pc}
\tablecaption{\label{table:observation}Cen A observation summary}
\tablehead{\colhead{Year} & \colhead{Month} & \colhead{Date} & \colhead{\shortstack{UT\\(hh:mm:ss)}} & \colhead{$u~ (\textrm{M}\lambda)$} & \colhead{$v~(\textrm{M}\lambda)$} & \colhead{\shortstack{SEFD\tablenotemark{a}\\(APEX, Jy)}} & \colhead{\shortstack{SEFD\tablenotemark{a}\\(SPT, Jy)}} & \colhead{\shortstack{Correlated Flux Density\tablenotemark{b}\\(Jy)}} & \colhead{SNR\tablenotemark{c}}}
\startdata
2015 & 1 & 17 & 07:20:00 & $-$2857 & 4125 & 7380 & 8560 & 0.45 $\pm$ 0.04 & 36 \\
2015 & 1 & 17 & 07:30:00 & $-$2720 & 4208 & 7250 & 8560 & 0.60 $\pm$ 0.05 & 69 \\
2015 & 1 & 17 & 07:40:00 & $-$2577 & 4288 & 7210 & 8560 & 0.58 $\pm$ 0.05 & 67 \\
2015 & 1 & 17 & 07:50:00 & $-$2430 & 4362 & 7200 & 8560 & 0.59 $\pm$ 0.05 & 68 \\
2015 & 1 & 17 & 08:20:00 & $-$1961 & 4560 & 6950 & 8560 & 0.56 $\pm$ 0.05 & 66 \\
2015 & 1 & 17 & 08:30:00 & $-$1797 & 4616 & 6960 & 8560 & 0.56 $\pm$ 0.05 & 66 \\
2015 & 1 & 17 & 08:40:00 & $-$1629 & 4667 & 6990 & 8560 & 0.47 $\pm$ 0.04 & 55 \\
2015 & 1 & 17 & 08:50:00 & $-$1459 & 4713 & 6990 & 8560 & 0.49 $\pm$ 0.04 & 57 \\
\enddata
\tablenotetext{a}{See Section~\ref{subsec:cal}.}
\tablenotetext{b}{Includes only statistical errors.}
\tablenotetext{c}{The first scan shows much lower SNR than the rest of scans. See Section~\ref{subsec:variability}.}
\end{deluxetable*}

On January 17, 2015, the SPT and APEX performed VLBI observations of several sources, including J0522-363, B1244-255, W Hya, Sagittarius A*, and Cen A. It was the first VLBI observation using the 1.4 mm SPT VLBI receiver. APEX used its 1.4 mm SHeFI receiver \citep{Belitsky:2007iv, 2008A&A...490.1157V}. APEX had previously demonstrated millimeter VLBI capability in an experiment with the SMT in Arizona \citep{2015A&A...581A..32W}.

\begin{figure}[b]
\begin{center}
\includegraphics[width=8.5cm]{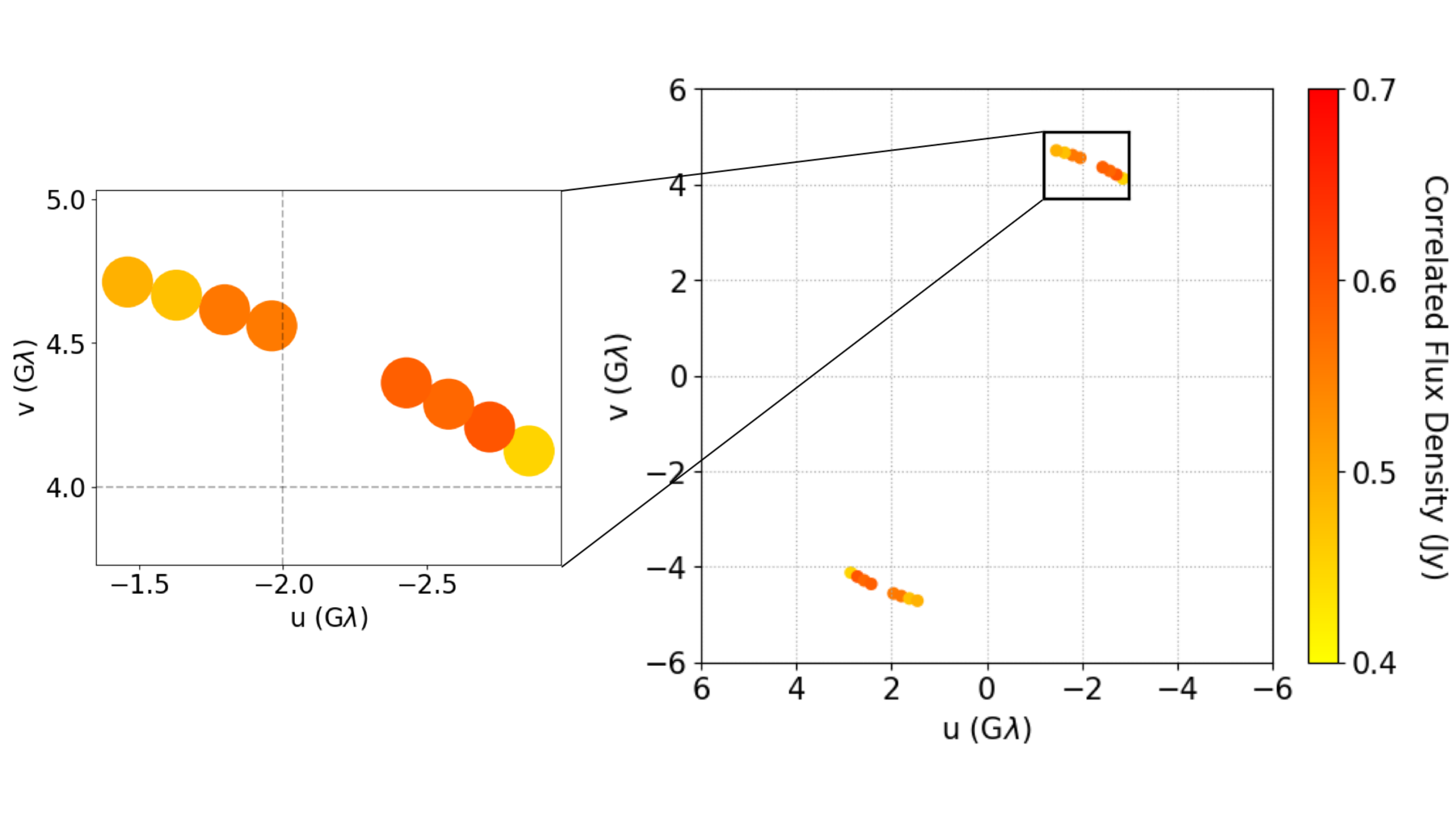} 
\end{center}
\caption{$(u, v)$ coverage of the SPT-APEX VLBI observation of Cen A at 1.4 mm. The color of the marker shows the strength of the correlated flux density at each $(u, v)$ coordinate and the size of the markers is irrelevant. Left is the expanded view around the coordinates.}
\label{fig:uv_coverage}
\end{figure}

The observation of Cen A included eight scans between 07:20 UT and 08:55 UT. Each scan was a 5-minute integration. Data were recorded for frequencies between 214.138 and 216.122 GHz, a receiving bandwidth of \mbox{2 GHz}, centered on 215.13 GHz, or 1.39 mm, in left circular polarization. At each station, the data were digitized by a ROACH2 (Reconfigurable Open Architecture Computing Hardware) digital backend \cite[R2DBE;][]{2015PASP..127.1226V} and recorded at 2-bit precision on Mark 6 recorders \citep{2013PASP..125..196W}. We correlated the data on the DiFX correlator \citep{2011PASP..123..275D} at the MIT Haystack Observatory. The fringe fitting of the correlated data was done by \texttt{fourfit} of Haystack Observatory Post-processing System\footnote{http://www.haystack.mit.edu/tech/vlbi/hops.html} (HOPS). We then segmented the data down to 1 s and incoherently averaged to produce the fringe amplitude \citep{1995AJ....109.1391R}. We found strong detections for all eight scans with signal-to-noise ratios (SNRs) between 36 and 69. The $(u, v)$ coverage of the scans is shown in Figure~\ref{fig:uv_coverage}. All scans correspond to a baseline length of approximately 5 G$\lambda$.

\subsection{Calibration}
\label{subsec:cal}
Visibility amplitude calibration is required to convert the correlated data to flux density units.  The system equivalent flux density (SEFD), the system temperature of the telescope divided by the gain, provides this calibration.  

The absolute calibration of APEX was determined from observations of Saturn. System temperatures were measured by observing an ambient temperature load and recorded during every scan. The receiver noise component of the system temperature was evaluated using an absorber cooled to $\sim$73~K. The SEFD uncertainty for APEX data is 7\%, based on the quadrature sum of the 5\% scatter between calibration measurements and the additional 5\% uncertainty in the absolute calibration scale from Saturn.

The absolute calibration of the SPT was determined from observations of Saturn and Venus. In 2015, the system temperature was not continuously monitored but was estimated from the combination of ambient temperature load observations on the day after the VLBI experiment and data from a 350~$\mu$m tipping radiometer on site \citep{2016PASP..128g5001R}. There are several factors that contribute to the uncertainty in the SEFD calibration. The lack of contemporaneous system temperature measurements during the observation contributes an uncertainty of 10\%, allowing for 25\% changes in the opacity between days. The translation between the VLBI signal chain and a separate power monitoring signal chain that was used for the planetary calibration observations contributes an additional 7\% uncertainty, inferred from the scatter between repeated measurements. The observed pointing drift during the observation from repeated pointing measurements suggests that there are possible 10\% changes in the telescope gain due to mis-pointing. The uncertainty due to the planet model is taken to be 5\%. In quadrature, these sum to 16\% uncertainty in the SEFD.

The mean SEFD for the Cen A scans is 7,100~Jy for APEX, while the SPT SEFD is fixed at 8,560~Jy. Table~\ref{table:observation} summarizes the scans, including baseline lengths, SEFDs, visibility amplitudes in flux density units, and the detection SNRs.


\section{Data Analysis}
\label{sec:analysis}

VLBI imaging of Cen A at wavelengths longer than 1.3 cm reveals that its inner jet structure has a number of jet components emerging from a
bright core \citep{1998AJ....115..960T, 2001AJ....122.1697T, 2006PASJ...58..211H, 2011A&A...530L..11M, 2014A&A...569A.115M}. \cite{2001ApJ...546..210T} compared the VLBA observation images at 13.6, 6.0, and 3.6 cm to estimate the spectral index around the subparsec-scale jet of Cen A. They reported that the spectrum towards the nucleus was highly inverted (increasing flux density with decreasing wavelength), and the core began to dominate the jet at 3.6 cm. \cite{2001AJ....122.1697T} also discovered that the jet components present in the 3.6 cm images were not observed in 1.3 cm images and only the core component was detected.

\cite{2011A&A...530L..11M} used a southern VLBI array at wavelengths of 1.3 and 3.6 cm to produce images of Cen A with higher resolution and image fidelity. They also found that the core region is brighter at 1.3 cm than at 3.6 cm, while the spectral index along the jet steepens away from the core, with jet components dimmer at 1.3~cm than 3.6~cm. They modeled the innermost region of the jet with two Gaussian components at \mbox{1.3 cm}. Furthermore, the 7 mm data of \cite{1997ApJ...475L..93K} implied that the structure is close to a single resolved component, although they were not able to form an image due to limited $(u, v)$ coverage. These observations were made over the past two decades, and the structure observed in previous epochs may differ from what would be measured now in this dynamic source. However, extrapolating this general trend with wavelength, we expect the 1.4 mm emission to be dominated by the optically thick core region. We model the core as a single circularly symmetric Gaussian component. This model is a typical choice for VLBI observations with poor $(u, v)$ coverage \cite[e.g.,][]{1997ApJ...475L..93K}. Because the SPT-APEX baseline rotated little in the $(u, v)$ plane during our observation (Figure~\ref{fig:uv_coverage}), we have little power to constrain an ellipticity parameter for the source model. Similarly, our observations do not constrain source structure, though from the considerations above we believe the approximation of a single source is plausible. The orientation of the effectively one-dimensional $(u,v)$ coverage is minimally sensitive to the elongation of the jet, so our size constraints primarily pertain to the perpendicular direction, though we have assumed circular symmetry in the source. The following discussion should be considered with the limitations of our data in mind, including the discussion of the observed visibility amplitude variability in Section~\ref{subsec:variability}.
\subsection{Brightness Temperature}
\begin{figure}[b]
\begin{center}
\includegraphics[width=8.5cm]{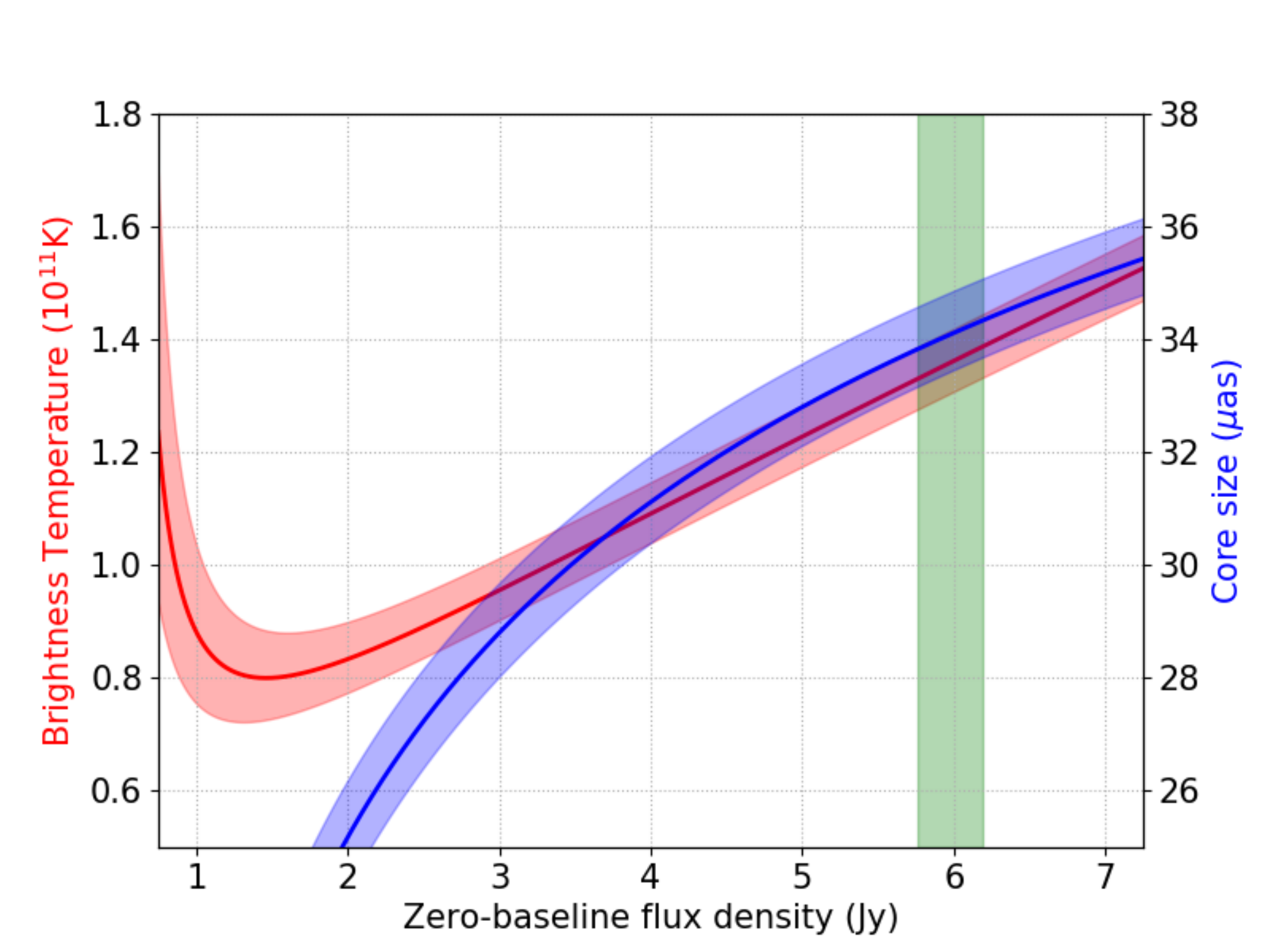} 
\end{center}
\caption{Ranges of the brightness temperature and the size of the Cen A core region versus the true zero-baseline flux density. The shaded regions around the red and blue curves indicate the 1$\sigma$ calibration uncertainties. The flux density measured by the SMA, $V_0$ = 6.0 $\pm~0.2$~Jy, is marked with the vertical green band.}
\label{fig:tblimit_coresize}
\end{figure}

\begin{figure}[b]
\begin{center}
\includegraphics[width=8.5cm]{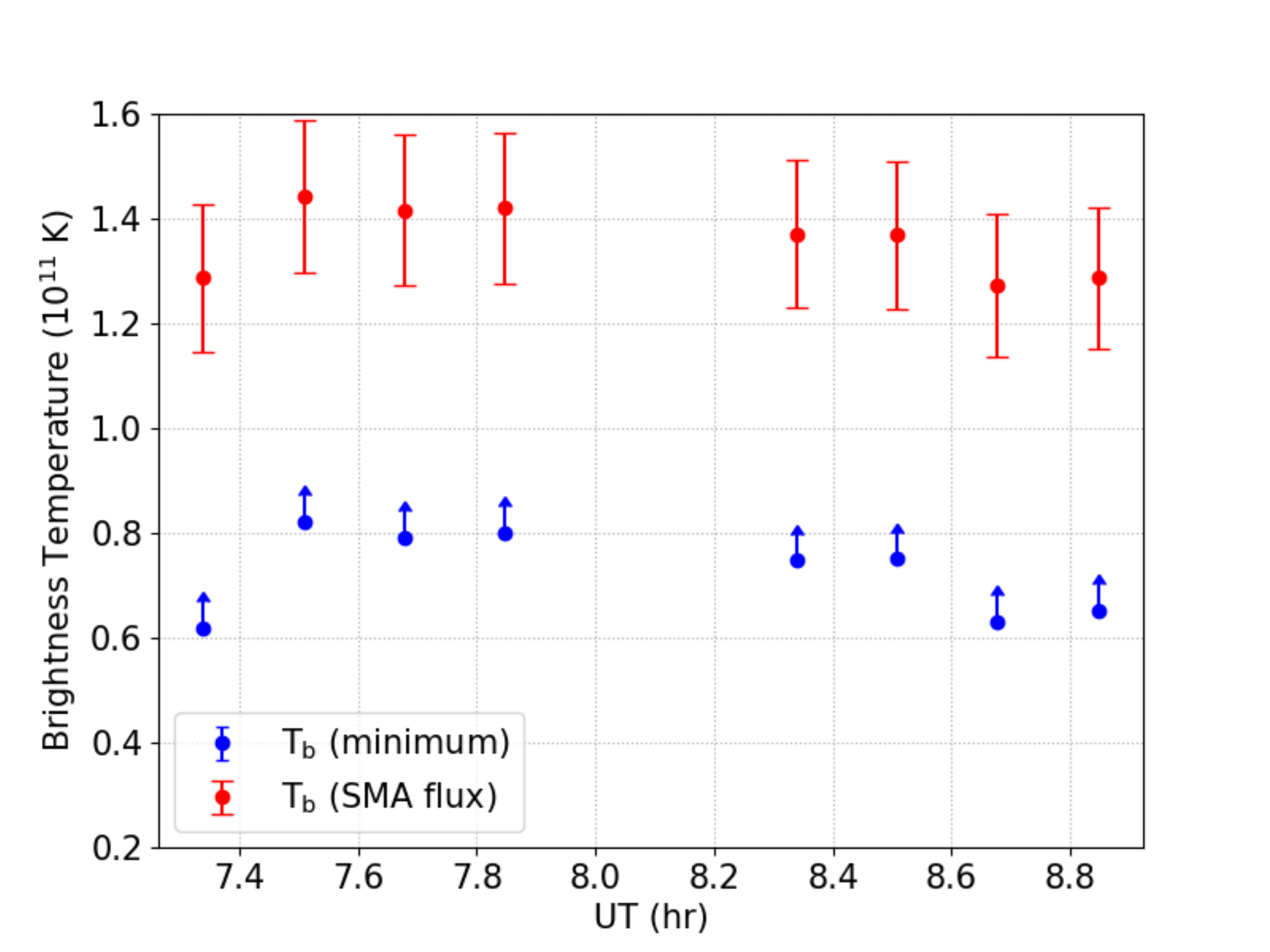} 
\end{center}
\caption{Brightness temperatures of the Cen A core as a function of UT on January 17, 2015. The lower limit using only the SPT-APEX baseline correlated flux density (blue) and the brightness temperature derived with additional SMA zero-baseline data (red) are shown.}
\label{fig:brightnesstemp}
\end{figure}

The brightness temperature of a circular Gaussian source of total (zero-baseline) flux density $V_0$ and full-width at half maximum (FWHM) $\phi$ observed at frequency $\nu$ is given by
\begin{equation}
T_\textrm{b} = 2\ln{2} \frac{c^2}{{k_B} {\nu}^2} \frac{V_0}{\pi {\phi}^2},
\end{equation}
where $k_B$ is the Boltzmann constant and $c$ is the speed of light. For correlated flux density $V_q$ measured on baseline length $B$, the implied FWHM is
\begin{equation}
\phi = \frac{2 \sqrt{\ln{2}}}{\pi} \frac{1}{B} \frac{c}{{\nu}} \sqrt{\ln{\left(\frac{V_0}{V_q}\right)}}
\label{eq:phi}
\end{equation}
and the brightness temperature is
\begin{equation}
T_{\textrm{b}} = \frac{\pi}{2k_B} \frac{B^2 V_0}{\ln(V_0 / V_q)}.
\label{eq:tb}
\end{equation}
We have no measurement of the zero-spacing flux density that was obtained at the same time as our VLBI observation, so deriving a brightness temperature from Equation~\ref{eq:tb} requires that we fix $V_0$ to values obtained at other times. Taking the mean observed correlated flux density, $V_q = 0.54 \pm 0.05$, the derived brightness temperature varies by a factor of roughly 2 depending on the choice of $V_0$ for reasonable values, as shown in Figure~\ref{fig:tblimit_coresize}. Without making any assumption about $V_0$ we can derive the minimum brightness temperature \citep{2015A&A...574A..84L} by setting $\partial T_b / \partial V_0 = 0$ in Equation~(\ref{eq:tb}), resulting in 
\begin{equation}
T_{\textrm{b, min}} = \frac{\pi e}{2k_B} B^2 V_q.
\end{equation}
Figure~\ref{fig:brightnesstemp} shows the minimum brightness temperature as a function of UT, and $T_{\textrm{b, min}}$ from our observation is roughly $7 \times 10^{10}$~K.

We can narrow the range of plausible values for $V_0$ by looking for contemporaneous measurements of the flux density of Cen A. The SMA monitors\footnote{http://sma1.sma.hawaii.edu/callist/callist.html} the flux density of radio sources for use as gain calibrators and secondary flux standards at mm wavelengths \citep{2007ASPC..375..234G}, and has two observations at 1 mm within a week of our measurement:
\begin{itemize}
\setlength\itemsep{0em}
\item[] 5.9 $\pm$ 0.3 Jy (January 16, 2015) and
\item[] 6.1 $\pm$ 0.3 Jy (January 22, 2015).
\end{itemize}
The SMA has sub-arcsecond resolution, and the observations are nearly contemporaneous. The SMA data therefore provide a useful total flux density (or “zero-baseline flux density”) for the AGN component that should not resolve out any of the 1.4~mm emission from the AGN core but that will spatially filter the emission from the dust and star formation of NGC 5128. Unless there are distant jet hot spots at 1 mm that are not predicted by the spectra of knots seen at longer wavelengths and not seen in other arcsecond-resolution 1~mm images \citep[e.g.,][]{2017ApJ...851...76M}, the SMA flux should be dominated by the core. Adopting $V_0=6.0 \pm 0.2$ Jy, the mean of the two SMA flux densities (marked by the green vertical band in Figure~\ref{fig:tblimit_coresize}), the brightness temperature implied for the Cen A core is $(1.4 \pm 0.2) \times 10^{11}~\textrm{K}$.

Table~\ref{table:brightnesstemp} lists the brightness temperatures measured for Cen A at other wavelengths. Included in this table are simultaneous measurements from 19.0 cm to 7~mm made with the VLBA in 2013 \citep[project code BH182B;][]{2013EPJWC..6108004H}, which have not previously been published. Our 1.4 mm brightness lower limit is comparable to that seen at other wavelengths, while the brightness temperature estimated including the SMA zero-spacing flux density is higher than nearly all others. This is what would be expected if our observation is sensitive to emission deeper in the synchrotron core due to decreasing synchrotron optical depth at shorter wavelengths. The estimated brightness temperature is still below the $\sim10^{12}$~K inverse Compton limit \citep{1969ApJ...155L..71K} even at this shortest wavelength, and close to the equipartition limit of $\sim10^{11}$~K \citep{1994ApJ...426...51R}. Doppler boosting from relativistic motion of the jet does not appear to be important for this source \citep[$\delta \sim 1$;][]{1998AJ....115..960T, 2007A&A...471..453M, 2014A&A...569A.115M}, so the brightness temperature does indicate that the region is near equipartition and it is likely that the amount of energy stored in particles and the magnetic field are similar.

\begin{deluxetable*}{cccccc}
\tablecolumns{6}
\tablewidth{0pc}
\tablecaption{Brightness temperature, flux density and the size of Cen A core region}
\tablehead{\colhead{Wavelength (mm)} & \colhead{Frequency (GHz)} & \colhead{Brightness temperature (K)} & \colhead{Flux density (Jy)\tablenotemark{a}} & \colhead{Size ($\mu$as)} & \colhead{Reference}}
\startdata
190 & 1.6 & $4.6 \times 10^{9}$ & 0.83 &11000 $\pm$ 530 & VLBA, March 2013\\
130 & 2.3 & $2.4 \times 10^{9}$ & 1.02 & 11000 $\pm$ 540 & VLBA, March 2013\\
61 & 4.9 & $5.7 \times 10^{8}$ & 0.69 & 7900 $\pm$ 400 & VLBA, March 2013\\
61 & 4.9 & $2.2 \times 10^{10}$\tablenotemark{b} & 0.30 & 2000 $\pm$ 470 & \cite{2006PASJ...58..211H}\\
36 & 8.4 & $4.7 \times 10^{9}$ & 3.26 & 3600 $\pm$ 180 & VLBA, March 2013\\
36 & 8.4 & $5.9 \times 10^{9}$\tablenotemark{c} & 2.47 & 2700 $\pm$ 470 & \cite{1998AJ....115..960T}\\
36 & 8.4 & $2.1 \times 10^{9}$\tablenotemark{c} & 2.25 & 4300 $\pm$ 840 & \cite{2001AJ....122.1697T}\\
36 & 8.4 & $1.5 \times 10^{11}$\tablenotemark{b} & 0.53 & 270 $\pm$ 60 & {\cite{2011A\string&A...530L..11M}}\\
36 & 8.4 & $6.5 \times 10^{10}$\tablenotemark{d} & 1.09 & 580 $\pm$ 160 & {\cite{2014A\string&A...569A.115M}}\\
24 & 12 & $3.1 \times 10^{9}$ & 2.07 & 2400 $\pm$ 120 & VLBA, March 2013\\
20 & 15 & $4.9 \times 10^{9}$ & 3.15 & 1900 $\pm$ 94 & VLBA, March 2013\\
14 & 22 & $4.0 \times 10^{9}$\tablenotemark{c} & 2.35 & 1200 $\pm$ 170 & \cite{2001AJ....122.1697T}\\
13 & 22 & $4.0 \times 10^{9}$ & 2.34 & 1300 $\pm$ 130 & VLBA, March 2013\\
13 & 22 & $3.0 \times 10^{10}$\tablenotemark{b} & 1.21 & 680 $\pm$ 90 & {\cite{2011A\string&A...530L..11M}}\\
7 & 43 & $1 \times 10^{10}$ & 3.15 & 500 $\pm$ 100 & {\cite{1997ApJ...475L..93K}} \\
7 & 43 & $4.5 \times 10^{9}$ & 2.17 & 570 $\pm$ 57 & VLBA, March 2013\\
\multirow{2}{*}{1.4} & \multirow{2}{*}{215} & $\geq 7 \times 10^{10}$\tablenotemark{e} & \multirow{2}{*}{5.98} & \multirow{2}{*}{34.0 $\pm$ 1.8} & \multirow{2}{*}{This work} \\
& & $1.4 \times 10^{11}$\tablenotemark{f} & 
\enddata
\tablenotetext{a}{The flux density uncertainty given in the literature was incorporated to derive the size uncertainty. When there were multiple measurements, we considered the standard deviation as an error. For the 2013 VLBA data, we assumed the calibration errors of \mbox{10~\%} for 7 and 1.4~mm, and \mbox{5~\%} for wavelengths longer than 7~mm.}
\tablenotetext{b}{Brightness temperature of the highest flux density component.}
\tablenotetext{c}{Brightness temperature calculated using mean flux density and the size of the core during the observation period.}
\tablenotetext{d}{Mean of the brightness temperatures of the core region, derived during the seven epochs of observations.}
\tablenotetext{e}{Lower limit from a single baseline measurement.}
\tablenotetext{f}{Incorporating SMA flux density in addition to the single baseline measurement.}
\vspace{0pt}
\label{table:brightnesstemp}
\end{deluxetable*}

\subsection{Core Size}
Assuming that the 1.4~mm emission primarily arises from a single circularly symmetric Gaussian component, we can estimate the size using the zero-baseline flux density. If we again use $V_0=6.0$ Jy, as measured by the SMA, and the uncertainties in the SMA flux density as well as the correlated flux density, the FWHM size is $\phi = 34.0 \pm 1.8~\mu\textrm{as}$ from Equation~(\ref{eq:phi}). This corresponds to $\sim$120~$R_{\textrm{sch}}$ or \mbox{$\sim$0.7 light-day}, although the mass of the black hole is uncertain at the 50\% level, and $R_{\textrm{sch}}$ scales linearly with the mass. Figure~\ref{fig:tblimit_coresize} plots the range of core sizes as a function of zero-baseline flux density. Because the SMA flux density may include a contribution from outside the core \citep{2008A&A...490...77W}, the size we calculated needs to be considered as an upper limit.

\begin{figure}[b]
\begin{center}
\includegraphics[width=8.5cm]{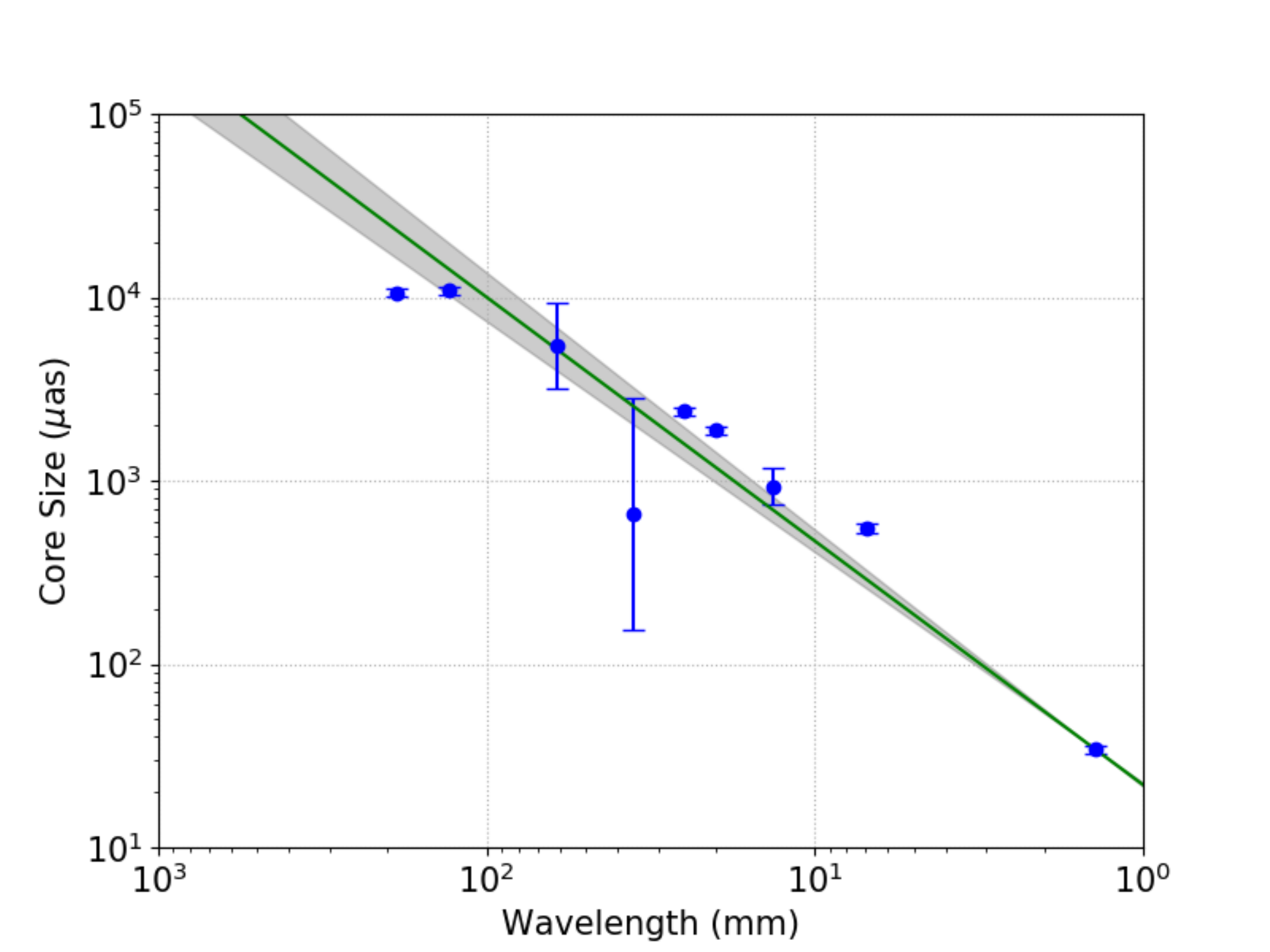} 
\end{center}
\caption{Size of the core region in 19.0 cm, 13.0 cm, \mbox{6.1 cm}, 3.6 cm, 2.4 cm, 2.0 cm, 1.3 cm, 7 mm, and 1.4 mm from the references in Table~\ref{table:brightnesstemp}. We use weighted average of multiple observation results, for each wavelength. Green line is the best fit to the data with the index $\alpha = 1.3 \pm 0.1$, where the core size $\phi \propto {\nu}^{-\alpha}$. We use 1.4 mm data as an intercept point of the fit.}
\label{fig:coresize}
\end{figure}

Figure~\ref{fig:coresize} compiles the data from the literature as well as the archival VLBA data in Table~\ref{table:brightnesstemp}. We used the flux density and the size of the core when the component identification is available in the original papers. \cite{2006PASJ...58..211H} and \cite{2011A&A...530L..11M} employed multiple components to model the core region, and the core size in Table~\ref{table:brightnesstemp} employs the mean of those component sizes. We adopted parameters for the brightest component when no other information is available (see the notes in Table~\ref{table:brightnesstemp}). The core size decreases with the increasing observing frequency (decreasing observing wavelength) and we fit the data to infer its frequency dependence. The size of the radio core should correspond to the region within which the synchrotron optical depth exceeds unity \citep{1979ApJ...232...34B, 1981ApJ...243..700K}. Referencing the size evolution with wavelength to the current 215~GHz (1.4~mm) results as
\begin{equation}
\frac{\phi}{\phi_{\textrm{215 GHz}}} = {\left(\frac{\nu}{\textrm{215 GHz}}\right)}^{-\alpha},
\end{equation}
where $\nu$ is the observing frequency, the best fit gives $\alpha = 1.3 \pm 0.1$. The dependence is similar to the angular size-frequency relation found in parsec-scale jets \citep{2008arXiv0811.2926Y} and the core shift of a jet observed in M87 \citep{2011Natur.477..185H}. For a conical jet, the shift and the change in size are linearly proportional to each other, following the same frequency dependence.

\subsection{Spectrum of the Core}

\begin{figure}
\begin{center}
\includegraphics[width=8.5cm]{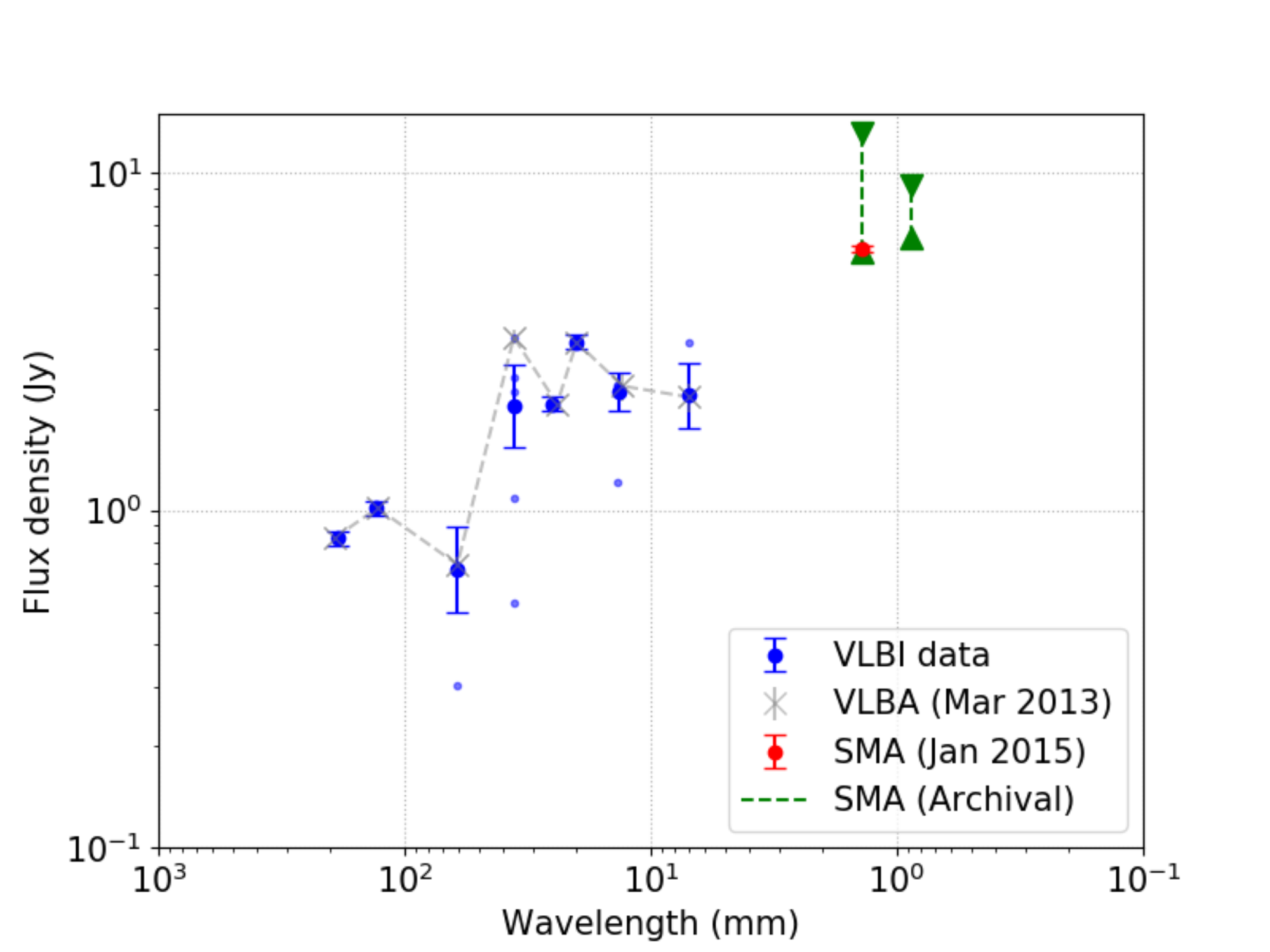} 
\end{center}
\caption{Spectrum of the Cen A core region. The VLBI observations between 19.0 cm and 7 mm (blue), the SMA 1.3 mm data of January 2015 (red), and the range of 1.3 mm and 0.8 mm SMA archival data (green) are plotted. The VLBI data use weighted average of multiple observations at each wavelength. The markers of the SMA archival data indicate minimum and maximum flux densities during the 12 years of observing period. The grey dotted line shows the simultaneous VLBA flux measurement in March 2013, and the small blue dots are the observations used to estimate the average spectrum.}
\label{fig:core_spectrum}
\end{figure}

The high brightness temperature of the core, greater than $10^{11}$~K, and the frequency dependence of its size 
are suggestive of wavelength-dependent synchrotron self-absorption. Flux-density measurements from the literature using interferometric arrays are presented in Figure~\ref{fig:core_spectrum}, including VLBI measurements of the core flux density between 19.0~cm and 7~mm (3.6 $\times$ 1.2 milliarcsec beam size at 7~mm), arcsecond-resolution measurements at high frequency from the SMA used to set the zero-baseline flux density. The flux density range observed by the SMA over 12 years of monitoring is also shown. The spectrum of the \mbox{Cen A} core increases until $\sim$3 mm, and the single-dish data in \cite{1993MNRAS.260..844H} show that the core has relatively flat spectrum shortward of 2 mm (Figure~1 of \citealt{1997ApJ...475L..93K} and Figure~5 of \citealt{2010ApJ...719.1433A}). 

The VLBI data show the core flux density  increasing with decreasing wavelength, a trend that continues to less than 1 mm if the SMA data also trace the core emission. Previous analyses, based on single-dish data with significantly lower resolution \cite[tens to hundreds of arcseconds, e.g.,][]{2007A&A...471..453M, 2008A&A...483..741I} show a spectrum that decreases in flux density with decreasing wavelength. At centimeter wavelengths these spectra are clearly dominated by the extended radio jet emission that fades most quickly toward short wavelengths because of synchrotron cooling. Because the compilation of Figure~\ref{fig:core_spectrum} and Table~\ref{table:brightnesstemp} selects the bright central components of VLBI images (in most cases), it provides the most applicable comparison for the 1.4 mm data presented here. The core flux density spectrum between 19.0 cm and 1.4 mm follows \mbox{$S_\nu  \propto \nu ^{0.39 \pm 0.07}$}. Of course, the flux density measurements span more than 20 years and show substantial variability (small points in Figure~\ref{fig:core_spectrum}), even when obtained at many wavelengths at once (grey dotted line in Figure~\ref{fig:core_spectrum}), so the spectral index can only be considered as a coarse average value. Nevertheless, the spectrum appears to be inverted, which can be produced by an optically thick, non-uniform synchrotron source \citep{1976A&A....52..439D}.

\subsection{Variability}
\label{subsec:variability}

The measured flux density as a function of UT is shown in Table~\ref{table:observation}. The flux density fluctuates from 0.45~Jy to 0.60~Jy over 1.5~hours, a variation of 16\% from the mean value, with the most significant deviation found in the first scan. The most likely explanation for variations between the scans after the first is a combination of calibration errors due to pointing shifts during SPT commissioning and atmospheric decorrelation within the scans. We note that the first scan is missing roughly 50\% of its data, which suggests that there may be further undiagnosed problems that reduce the amplitude of the correlation.
 
 The baseline length changes very little over the course of these observations (2\%), which, for a circularly symmetric Gaussian source, would lead to much less variation than we observe (8\%). If the 1.4~mm source is actually elliptical or composed of multiple components, the variation in visibility amplitude along the 1.5~G$\lambda$-long arc traced by the baseline in the ($u,v$) plane could be larger. If we assume that the visibility variation is induced by ellipticity in the Gaussian source, at a position angle aligned with the center of the $(u, v)$ track, the best-fit axis ratio would be \mbox{1.6 : 1}.

The nucleus of Cen A is known to be variable on daily to yearly time-scales at different wavelengths \citep{1971ApJ...170L..11W, 1974ApJ...194L.135K, 1989AJ.....98...27M, 1993MNRAS.264..807B, 2008A&A...483..741I, 2014A&A...569A.115M}. The light crossing time of the core limits the variability to $\sim$ 1 day, though Doppler effects can shorten this time scale for beamed sources.


\section{Conclusion}

The first VLBI observations from the South Pole Telescope have detected correlated emission on a 7000~km, 5~G$\lambda$ baseline to the APEX telescope. With these data, we constrain the brightness temperature of the Cen A core region at 40~$\mu$as resolution. The calculated core size is 120~$R_\textrm{sch}$ for the $5.5 \times 10^7 M_{\odot}$ central black hole. The frequency dependence of the core size and its spectrum suggest that we are detecting the self-absorbed synchrotron emission region around the black hole. Once the other stations participate, the full EHT array will yield significantly better, two-dimensional, $(u, v)$ coverage, resolution, and sensitivity, allowing imaging of the Cen A core and more detailed investigation of this source.

\acknowledgments
J.K., D.P.M., and S.S.D acknowledge support from NSF grants AST-1207752 and AST-1440254. S.S.D. acknowledges support for this work from NSF under grants AST-1207704 and  AST-1310896. J.W.H. acknowledges support from NSF grant AST-1402161. C.M. acknowledges support from the ERC Synergy Grant ``BlackHoleCam: Imaging the Event Horizon of Black Holes'' (Grant 610058). E.R. acknowledges partial support from the MINECO grants AYA-2012-38491-C02-01, AYA2015-63939-C2-2-P, and the Generalitat Valenciana grant PROMETEOII/2014/057. The South Pole Telescope program is supported by the National Science Foundation through grant PLR-1248097. Partial support is also provided by the NSF Physics Frontier Center grant PHY-0114422 to the Kavli Institute of Cosmological Physics at the University of Chicago, the Kavli Foundation, and the Gordon and Betty Moore Foundation through Grant GBMF\#947 to the University of Chicago. The Submillimeter Array is a joint project between the Smithsonian Astrophysical Observatory and the Academia Sinica Institute of Astronomy and Astrophysics and is funded by the Smithsonian Institution and the Academia Sinica. We thank Chris Kendall and Dave Pernic for their assistance during the receiver installation at the South Pole.
This research made use of \texttt{Astropy}, a community-developed core Python package for Astronomy \citep{2013A&A...558A..33A}.

\software{Astropy (The Astropy Collaboration 2013), HOPS (http://www.haystack.mit.edu/tech\\/vlbi/hops.html)}

\bibliographystyle{aasjournal}
\bibliography{ms.bib} 

\end{document}